\documentclass[12pt]{article}

 \usepackage{amsmath,amsthm}
 \usepackage{amsfonts,amscd,amssymb}
 \usepackage[matrix,arrow,curve]{xy}
 \usepackage{rotating,array}
 \usepackage[english]{babel}

 \theoremstyle{plain}
 \newtheorem{theorem}{Theorem}

\theoremstyle{definition}

\begin{document}

\title{SEIBERG--WITTEN THEORY AS A COMPLEX VERSION OF ABELIAN HIGGS MODEL}
\author{Armen SERGEEV\footnote{While preparing it the author was
partially supported by the RFBR grants 16-01-00117, 13-02-91330 and
Scientific Program of Presidium of RAS "Nonlinear Dynamics"}}

\date{Steklov Mathematical Institute, Moscow}

\maketitle
\bigskip

\hfill To the memory of Professor Lu Qikeng

\section*{}

I first met Professor Lu Qikeng in the Mittag-Leffler Institute in
Stockholm in 1988 during the Year of Several Complex Variables
organized by John-Eric Fornaess and Christer Kiselman. When I
entered the beautiful building of this institute placed in Stockholm
suburbs the first men whom I met were Professors Lu and Kiselman
sitting under the portrait of Sofya Kovalevskaya. After I was
introduced to Professor Lu we started to discuss the topics of our
mutual mathematical interest. At that time it was complex analysis
in matrix spaces --- a traditional theme for Chinese mathematicians
starting from Professor Hua Lookeng. I was looking for a
generalization of the well known criterion of holomorphic convexity
of $n$-circled (or Reinhardt) domains to the matrix case. When I
told Professor Lu about this problem he decided to propose it to his
student Zhou Xiangyu. Shortly after that Zhou Xiangyu has solved the
problem (it was solved independently also by Eric Bedford and Jiri
Dadok).

It was the beginning of our collaboration and friendship with
Professor Lu and Zhou Xiangyu. In 90's (I do not remember the year
precisely) Professor Lu has visited Steklov Institute where he met
my teacher Professor V.S.Vladimirov who told him about a
long-standing conjecture in axiomatic quantum field theory. It was
the so called "extended future tube conjecture" posed by A.S.
Wightman. V.S.Vladimirov was a great enthusiast of this problem and
proposed it to all of his students working in quantum field theory
and several complex variables. Professor Lu became also interested
in this problem and invited me to come to Beijing and give a
mini-course on it in the Institute of Mathematics of Academia
Sinica. I came to Beijing on the eve of 1999 in the middle of a cold
and windy winter and gave three lectures on the extended future tube
conjecture presenting several reformulations of this problem. From
this time my relations with Professor Lu were mostly via Zhou
Xiangyu. He visited Steklov Institute several times almost for three
years in total and became the first Chinese mathematician who got
the Doctor of Science (second doctor) degree from this Institute for
the solution of the extended future tube conjecture and other
results on invariant domains of holomorphy.

This story shows how Professor Lu cared about his students and how
far-seeing he was as a mathematician. I should add that being a very
nice person we was easy to deal with. I always felt relaxed while
meeting and talking to him despite certain language problems. His
untimely death is a big loss for me personally and, I believe, for
all mathematics.

\section*{Introduction}

In their papers \cite{SW1},\cite{SW2} Seiberg and Witten have
proposed what is called now the Seiberg--Witten theory. Motivated by
this theory, Witten \cite{Wit} introduced the Seiberg--Witten
equations which were used to define a new kind of invariant of
smooth 4-dimensional manifolds. The Seiberg--Witten equations,
opposite to the Yang--Mills duality equations which are conformally
invariant, are not invariant under scale transformation so in order
to produce invariants from these equations one should plug the scale
parameter $\lambda$ into them and consider the scale limit
$\lambda\to\infty$.

If we consider such a limit in the case of 4-dimensional symplectic
manifolds a solution of Seiberg--Witten equations will concentrate
in a neighborhood of some pseudoholomorphic curve (more precisely, a
pseudoholomorphic divisor) while the equations reduce to families of
static Abelian Higgs equations defined in the normal planes to the
limiting pseudoholomorphic curve. Such a limit is called adiabatic
as well as the reduced Seiberg--Witten equations. Solutions of these
equations may be considered as families of static solutions of the
Abelian Higgs model in the complex plane with a complex parameter
$z$ running along the pseudoholomorphic curve. This parameter plays
the role of complex time while the reduced Seiberg--Witten equations
have the form of a nonlinear $\bar\partial$-equation with respect to
$z$.

It turns out that this construction has a non-trivial
$(2+1)$-dimensional analogue. Namely, if we consider in the
$(2+1)$-dimensional Abelian Higgs model the "slow-time" limit then
Abelian Higgs equations will reduce to the adiabatic equations with
solutions given by the geodesics on the moduli space of static
Abelian Higgs solutions (called otherwise vortices) in the metric
generated by the kinetic energy functional.

Thus we may consider the reduced Seiberg--Witten equations as a
$(2+2)$-dimensional analogue of the adiabatic equations in Abelian
Higgs case. Solutions of these equations can be treated as complex
analogues of adiabatic geodesics while the nonlinear
$\bar\partial$-equation may be considered as a complex analogue of
the Euler equation for these geodesics.

\section{Seiberg--Witten theory}
\label{sec1}

In this section we recall briefly some basic facts from
Seiberg--Witten theory(cf. \cite{Don},\cite{Sal},\cite{Ser} for a
more detailed presentation).

\subsection{Spinor algebra}
\label{ssec11}

\textbf{Clifford algebras}. Let $V=\mathbb R^n$ be an
$n$-dimensional Euclidean vector space provided with the Euclidean
metric and an orthonormal basis $\{e_i\}$, $i=1,\ldots,n$. The
\textit{Clifford algebra} $\text{Cl}(n)$ is generated by the
elements $1,e_1,e_2,\ldots,e_n$, satisfying the following relations:
$$
e_{i}^2=-1,\quad e_ie_j+e_je_i=0\ \text{for}\ i\neq j.
$$
As a real vector space, $\text{Cl}(n)$ has dimension $2^n$ with the
basis given by the elements of the form $1,e_I:=e_{i_1}e_{i_2}\ldots
e_{i_k}$ where $I=\{i_1,i_2,\ldots,i_k\}$ is a subset of
$\{1,2,\ldots,n\}$ consisting of indices $i_1<i_2<\ldots<i_k$ with
$k=1,2,\ldots,n$. We denote by
$\text{Cl}^c(n):=\text{Cl}(n)\otimes_{\mathbb R}\mathbb C$ the
complexification of this algebra.

\textbf{Spin group}. Denote by $\text{Pin}(n)$ the subgroup of the
multiplicative group $\text{Cl}^{*}(n)$ of the Clifford algebra,
generated by the unit vectors $v\in V$, i.e. by vectors $v$ with
$|v|=1$. The \textit{spinor group} $\text{Spin}(n)$ is the identity
connected component of $\text{Pin}(n)$. There is an exact sequence:
$$
\begin{CD}
0 @>>> \mathbb Z_2 @>>> \text{Spin}(n) @>\pi>> \text{SO}(n) @>>> 0 .
\end{CD}
$$

\textbf{Spin representation}. In the case of even $n=2m$ there is a
spinor representation of the group $\text{Spin}(2m)$ in the
($2^m$)-dimensional Hermitian complex vector space $W$:
$$
\Gamma:\text{Spin}(2m)\longrightarrow\text{End\,}W,
$$
which extends to the representation of the whole complexified
Clifford algebra
$$
\Gamma:\text{Cl}^c(V)\to\text{End}\,W.
$$
The action of $\text{Cl}^c (V)$ on $W$ is called the
\textit{Clifford multiplication} while elements of $W$ are called
the \textit{spinors}.

\textbf{Semi-spinor spaces}. Define the \textit{Clifford volume
element} $\omega$ by setting
$$
\omega:=e_1e_2\ldots e_{2m}\in\text{Cl}_{2m}(V).
$$
Then $\omega^2=(-1)^m$ so we can introduce the \textit{semispinor
spaces}
$$
W^{\pm}:=\{w\in W:\ \Gamma(\omega)w=\pm i^mw\}.
$$
Thus we obtain a decomposition $W=W^{+}\oplus W^{-}$ into the direct
sum of semi-spinor spaces which are interchanged under the Clifford
multiplication by vectors $v\in V$.

\textbf{Relation with exterior algebra}. Consider the exterior
algebra $\Lambda^{*}V$ of the space $V$. There is a linear
isomorphism
$$
\text{Alt}:\Lambda^{*}V\longrightarrow\text{Cl}(V)
$$
defined by associating with a form $v_1\wedge\ldots\wedge v_k$ the
element of $\text{Cl}(V)$ given by the alternating sum
$$
v_1\wedge\ldots\wedge v_k\longmapsto\frac{1}{k!}\sum_{\sigma\in
S_k}\text{sgn}(\sigma) v_{\sigma(1)}\cdot\ldots\cdot v_{\sigma
(k)}\,.
$$
By duality, we have also the isomorphism
$$
\text{Alt}':\Lambda^{*}(V')\to\text{Cl}(V')\cong\text{Cl}(V)
$$
where $V'$ is the (real) dual vector space of $V$.

Using the spin representation $\Gamma:\text{Cl}(V)\to \text{End}\,W
$, we can define
$$
\rho:=\Gamma\circ\text{Alt}':\Lambda^{*}(V')\to \text{End}\,W.
$$
The introduced map $\rho$ determines the Clifford multiplication by
forms from $\Lambda^*(V')$ in the space $W$. In particular, the
Clifford multiplication by 2-forms leaves the subspaces $W_{\pm}$
invariant. More precisely, the map $\rho$ associates with the
real-valued 2-forms the skew-symmetric traceless endomorphisms of
subspaces $W^{\pm}$, and with imaginary-valued 2-forms the Hermitian
traceless endomorphisms of these subspaces.

\textbf{Dimension 4}. If $\dim V =4$ then the subspace
$\Lambda^{2}(V')$ decomposes into the direct sum
$$
\Lambda^{2}(V')=\Lambda^{2}_{+}\oplus\Lambda^{2}_{-}
$$
of subspaces of selfdual and anti-selfdual forms with respect to
Hodge $*$-operator. In this case the map $\rho$ induces the
isomorphisms: $\Lambda^{2}_{\pm}\to\text{su}(W^{\pm})$ and
$$
\rho^{\pm}:\ \Lambda^{2}_{\pm}\otimes i\mathbb
R\longrightarrow\text{Herm}_0(W^{\pm}).
$$
The isomorphisms, inverse to $\rho^{\pm}$, are denoted by
$$
\sigma_{\pm}=(\rho_{\pm})^{-1}:\text{Herm}_{0}(W^{\pm})\to\Lambda^{2}_{\pm}
\otimes i\mathbb R.
$$

\textbf{Complex case}. In the case when $V=\mathbb C^n$ provided
with an Hermitian metric there exists a canonical spin
representation. The corresponding spinor space is given by
$$
W_{\text{can}}=\Lambda^{0,*}(V'):=\bigoplus_{q=0}^{n} \Lambda^{0,q}
(V').
$$
The spin representation $\Gamma_{\text{can}}$ on vectors $v\in V$ is
given by the formula
$$
\Gamma_{\text{can}}(v)w^{0,q}=v^{0,1}\wedge w^{0,q}-v^{1,0}\lrcorner
w^{0,q}
$$
where $v'=v^{1,0}+v^{0,1}$ is the dual covector of $v$ and
$w^{0,q}\in\Lambda^{0,q}(V')$. The semi-spinor spaces coincide with
$$
W_{\text{can}}^{+}=\Lambda^{0,\text{ev}}(V'),\quad
W_{\text{can}}^{-}=\Lambda^{0,\text{od}}(V').
$$

\textbf{$\text{Spin}^c$ group}. The group $\text{Spin}^c(n)$ is a
\textit{circle extension} of the group $\text{Spin}(n)$ defined as
$$
\text{Spin}^c(n)=\{z=e^{i\theta}x:\ x\in\text{Spin}(n), \theta\in
\mathbb R\}.
$$
There is an exact sequence
$$
\begin{CD}
0 @>>> \text{Spin}(V) @>>> \text{Spin}^c (V) @>\delta>> \text{U}(1)
@>>> 0
\end{CD}
$$
where $\delta:\,xe^{i\theta}\mapsto e^{2i\theta}.$ So
$\text{Spin}^c(V) = \text{Spin}(V)\times_{\mathbb Z_2}\text{U}(1)$.

\subsection{$\text{Spin}^c$-structures}
\label{ssec12}

\textbf{Definition}. Let $X$ be an oriented $n$-dimensional
Riemannian manifold and $P_{\text{SO}(n)}\to X$ is a principal
$\text{SO}(n)$-bundle of orthonormal bases on $X$. The
$\text{Spin}^c$-\textit{structure} on $P_{\text{SO}(n)}$ is an
extension of this bundle to a principal $\text{Spin}^c(n)$-bundle
$P_{\text{Spin}^c(n)}\to X$ together with a
$\text{Spin}^c$-invariant bundle epimorphism:
\begin{equation*}
 \begin{CD}
  P_{\text{Spin}^c(n)} @ >>> P_{\text{SO}(n)} \\
  @VVV  @VVV \\
  X  @=  X\\
 \end{CD}
\end{equation*}
where $\text{Spin}^c(n)$ acts on $P_{\text{SO}(n)}$ by the
homomorphism $\pi:\text{Spin}^c(n)\to\text{SO}(n)$.

\textbf{Characteristic bundle}. We can associate with the bundle
$P_{\text{Spin}^c(n)}$ the principal $\text{U}(1)$-bundle
$P_{\text{U}(1)}\to X$ so that the following diagram is commutative:
\begin{equation*}
\begin{CD}
P_{\text{Spin}^c(n)} @ >\delta>> P_{\text{U}(1)} \\
@VVV  @VVV \\
X  @=  X\\
\end{CD}
\end{equation*}
where $\text{Spin}^c(n)$ acts on $P_{\text{U}(1)}$ by the
homomorphism $\delta:\text{Spin}^c(n)\to\text{U}(1)$. The complex
line bundle $L\to X$, associated with $P_{\text{U}(1)}\to X$, is
called the \textit{characteristic bundle} of the given
$\text{Spin}^c$-structure, and its first Chern class $c_{1}(L)$ is
called the \textit{characteristic class} of the
$\text{Spin}^c$-structure.

\textbf{$\text{Spin}^c$-structures on vector bundles}. In a similar
way one can define a $\text{Spin}^c$-structure on an oriented
Riemannian vector bundle $V\to X$ of rank $n$, associated with a
principal bundle $P_{\text{SO}(n)}\to X$. The
$\text{Spin}^c$-\textit{structure} on $V\to X$ is the extension of
its structure group from $\text{SO}(n)$ to $\text{Spin}^c(n)$. In
other words, the bundle $V\to X$ \textit{admits} a
$\text{Spin}^c$-\textit{structure} if it is associated with a
principal $\text{Spin}^c(n)$-bundle $P_{\text{Spin}^c(n)}\to X$,
i.e. there exists a bundle isomorphism
$$
P_{\text{Spin}^c(n)}\times_{\text{Spin}^c(n)}\mathbb
R^n\longrightarrow V
$$
where $\text{Spin}^c(n)$ acts on $\mathbb R^n$ by the homomorphism
$\pi:\text{Spin}^c(n)\to\text{SO}(n)$.

In particular, one can take for such $V$ the tangent bundle $TX$. In
this case the $\text{Spin}^c$-structure on $TX$ is called the
$\text{Spin}^c$-\textit{structure on the manifold} $X$. Such a
structure exists on any 4-dimensional oriented Riemannian manifold
$X$.

\textbf{Definition in terms of spin representation}. In the case
when the rank of $V$ is even, i.e. $n=2m$, we can give an equivalent
definition of the $\text{Spin}^c$-structure on $V$ in terms of the
spin representation. Namely, a $\text{Spin}^c$-\textit{structure on
the bundle} $V$ of rank $2m$ is a pair $(W,\Gamma)$, consisting of a
complex Hermitian vector bundle $W\to X$ of rank $2^m$ (spinor
bundle) and a bundle homomorphism $\Gamma:V\to\text{End\,}W$ subject
to relations
$$
\Gamma^{*}(v)+\Gamma(v)=0,\quad
\Gamma^{*}(v)\Gamma(v)=|v|^2\,\text{Id}.
$$
Such a homomorphism extends to a bundle homomorphism
$\Gamma:\text{Cl}^c(V)\longrightarrow\text{End\,}W$ where
$\text{Cl}^c(V)$ is the bundle of complexified Clifford algebras,
associated with the oriented Riemannian vector bundle $V$. In
particular, $W$ can be decomposed into the direct sum $W=W^+\oplus
W^-$ of semi-spinor bundles. The characteristic line bundle of the
$\text{Spin}^c$-structure $(W,\Gamma)$ coincides with the bundle
$$
L_\Gamma:=P_{\text{Spin}^c(2m)}\times_{\text{Spin}^c(2m)}\mathbb
C\longrightarrow X
$$
where the action of the group $\text{Spin}^c(2m)$ on $\mathbb C$ is
given by the homomorphism
$\delta:\text{Spin}^c(2m)\longrightarrow\text{U}(1)$.

\textbf{Associated $\text{Spin}^c$-structures}. Suppose that an
oriented Riemannian vector bundle $V\to X$ of rank $2m$ has a
$\text{Spin}^c$-structure $(W,\Gamma)$. Then with any complex line
bundle $E\to X$ we can associate a new $\text{Spin}^c$-structure
$(W_E,\Gamma_E)$ by setting
$$
W_E:=W\otimes E,\quad \Gamma_E:=\Gamma\otimes\text{Id}.
$$
This new $\text{Spin}^c$-structure $(W_E,\Gamma_E)$ corresponds to
the principal $\text{Spin}^c(2m)$-bundle
$$
P_{\Gamma_E}=P_\Gamma\otimes_{\text{U}(1)}P_E
$$
where $P_\Gamma$ is the principal $\text{Spin}^c(2m)$-bundle,
associated with $(W,\Gamma)$ and $P_E$ is the principal
$\text{U}(1)$-bundle, associated with $E$. The characteristic bundle
of $\text{Spin}^c$-structure $(W_E,\Gamma_E)$ coincides with
$$
L_{\Gamma_E}:=L_\Gamma\otimes E^{\otimes 2}
$$
where $L_\Gamma$ is the characteristic bundle of $(W,\Gamma)$.

\textbf{Complex case}. In the case when $V\to X$ is a complex (or
almost complex) vector bundle of (complex) rank $n$, provided with a
complex (resp. almost complex) structure $J$, compatible with
Riemannian metric and orientation of $V$ we can construct a
\textit{canonical} $\text{Spin}^c$-\textit{structure}
$(W_{\text{can}},\Gamma_{\text{can}})$ on $V$. For this structure
$$
W_{\text{can}}:=\Lambda^{0,*}(V')
$$
where $V'$ is provided with the dual almost complex structure $J'$.
The Clifford multiplication map $\Gamma_{\text{can}}$ is given by
the same formula, as in the case of complex vector spaces. The
characteristic bundle $L_{\text{can}}$ coincides with the
anticanonical bundle $K'$ of $V$:
$$
K'=\Lambda^{0,n}(V').
$$

\subsection{$\text{Spin}^c$-connections and Dirac operator}
\label{ssec13}

\textbf{$\text{Spin}^c$-connection}. Suppose that $X$ is an oriented
Riemannian manifold of dimension $2m$ for which the tangent bundle
$TX$ can be provided with a $\text{Spin}^c$-structure $(W,\Gamma)$.
Denote by $\nabla_0$ the Levi-Civita connection on $TX$. Then the
\textit{$\text{Spin}^c$-connection} $\nabla$ on $X$ is an extension
of the Levi-Civita connection $\nabla_0$ to $W$. In other words, it
is a linear first order differential operator on the space
$C^\infty(X,W)$ of smooth sections of $W$ satisfying the following
Leibniz rule
$$
\nabla_u\left(\Gamma(v)s\right)=\Gamma(v)\nabla_us+\Gamma(\nabla_{0,u})s
$$
for any vector fields $u,v$ on $X$ and any smooth section $s$ of
$W$.

\textbf{$\text{Spin}^c$-connection form}. Denote by $\mathcal A$ the
connection form of the introduced $\text{Spin}^c$-connection
$\nabla$. It is a 1-form on the principal $\text{Spin}^c(2m)$-bundle
$P_\Gamma$ with values in the Lie algebra $\text{spin}^c(2m)$ of the
group $\text{Spin}^c(2m)$. This Lie algebra is equal to
$$
\text{spin}^c(2m)=\text{so}(2m)\oplus i\mathbb R
$$
where $\text{so}(2m)$ is the Lie algebra of the orthogonal group
$\text{SO}(2m)$. It follows from this representation that
$$
\mathcal A=\mathcal A_0+A
$$
where $\mathcal A_0$ is the connection form of the Levi-Civita
connection on $TX$ and $A$ is the trace part of the form $\mathcal
A$. The form $2A$ generates a connection on the characteristic
bundle $L_\Gamma$. In the case when $L_\Gamma$ has a square root,
i.e. a line bundle $L_\Gamma^{1/2}\to X$ such that
$L_\Gamma^{1/2}\otimes L_\Gamma^{1/2}=L_\Gamma$ (this condition is
fulfilled, e.g., for spin manifolds $X$) the form $A$ generates a
connection on $L_\Gamma^{1/2}$.

\textbf{Dirac operator}. Denote by $\nabla_A$ (resp. $d_A$) the
covariant derivative (resp. exterior covariant derivative) on
sections from $C^\infty(X,W)$ generated by the connection $\mathcal
A$. The \textit{Dirac operator} $D_A: C^\infty(X,W^+)\longrightarrow
C^\infty(X,W^-)$, associated with the connection $\mathcal A$, is
defined by the formula
$$
D_As=\sum_{j=1}^{2m}\Gamma(e_j)\nabla_{A,e_j}s
$$
where $s\in C^\infty(X,W)$ and $\{e_j\}$ is a local orthonormal
basis of $TX$. (This definition does not depend on the choice of
$\{e_j\}$.)

\textbf{Complex case}. If $X$ is a complex (or almost complex)
manifold of dimension $n$ then it can be provided with a canonical
$\text{Spin}^c$-structure $(W_{\text{can}},\Gamma_{\text{can}})$ and
associated canonical $\text{Spin}^c$-connection $\mathcal
A_{\text{can}}$. If $E\to X$ is a complex line bundle over $X$,
provided with a Hermitian connection $B$, then we can construct a
$\text{Spin}^c$-connection $\mathcal A_E$ on the associated
principal bundle $P_{\Gamma_E}$ by setting
$$
A_E:=A_{\text{can}}\otimes\text{Id}+\text{Id}\otimes B
$$
where $2A_{\text{can}}$ is the connection form of the canonical
connection on $L_{\text{can}}=\Lambda^{0,n}(T^*X)$. The
corresponding spinor space $W_E=\Lambda^{0,*}(X,E)$ is decomposed
into the direct sum
$$
W_E=W^+_E\oplus W^-_E
$$
with
$$
W_E^{+}=\Lambda^{0,\text{ev}}(X,E),\quad
W_E^{-}=\Lambda^{0,\text{od}}(X,E).
$$
The associated Dirac operator $D_{A_E}:
C^\infty(X,W_E^+)\longrightarrow C^\infty(X,W_E^-)$ coincides in
this case with the operator
$$
D_{A_E}=\bar\partial_B+\bar\partial_B^*
$$
where $\bar\partial_B$ is the covariant $\bar\partial$-operator and
$\bar\partial_B^*$ is the adjoint of $\bar\partial_B$.

\subsection{Seiberg--Witten equations on 4-dimensional Riemannian manifolds}
\label{ssec14}

\textbf{Seiberg--Witten equations}. Let $X$ be an oriented compact
Riemannian 4-manifold provided with a $\text{Spin}^c$-structure
$(W,\Gamma)$ and $\text{Spin}^c$-connection $\nabla_A$. Then the
associated \textit{Seiberg--Witten equations} (briefly:
SW-equations) have the form
$$
\left\{
\begin{aligned}
D_{A}\Phi &=0\\
F_{A}^{+}=&\ (\Phi\otimes\Phi^{*})_0
\end{aligned}\right.
$$
where $\Phi\in C^\infty(X,W^+)$ and
$(\Phi\otimes\Phi^{*})_0:=\Phi\otimes\Phi^{*}-\frac12|\Phi|^2\,\text{Id}
$ is the traceless Hermitian endomorphism associated with $\Phi$.
The term $F_{A}^{+}\in\Omega_+^2(X,i\mathbb R)$ is the selfdual part
of the curvature $F_A$.

The first SW-equation is the covariant Dirac equation. To explain
the meaning of the second SW-equation recall that for 4-dimensional
manifolds $X$ we have the decomposition
$$
\Lambda^2(T^*X)=\Lambda^2_+\oplus\Lambda^2_-
$$
of the bundle $\Lambda^2(T^*X)$ of 2-forms on $X$ into the direct
sum of subbundles $\Lambda^2_\pm\equiv\Lambda^2_{\pm}(T^*X)$ of
selfdual (resp. anti-selfdual) 2-forms with respect to Hodge
$\star$-operator. Then the Clifford multiplication determines an
isomorphism
$$
\sigma_+:\text{Herm}_0 (W^{+})
\longrightarrow\Omega_{+}^2(X,i\mathbb R)
$$
where $\Omega_{\pm}^2(X,i\mathbb R)$ is the space of sections of the
bundle $\Lambda^{2}_{\pm}(T^*X)\otimes i\mathbb R$ over $X$.

\textbf{Seiberg--Witten action functional}. The SW-equations are the
Euler--Lagrange equations for the following \textit{Seiberg--Witten
action} functional
$$
S(A,\Phi)=\frac12\int_X\left\{|F_A|^2+|\nabla_A\Phi|^2+
\frac{|\Phi|^2}{4}\left(s(g)+|\Phi|^2\right)\right\}\text{vol}(g)
$$
where $s(g)$ is the scalar curvature of $(X,g)$ and $\text{vol}(g)$
is its volume form.

The Seiberg--Witten equations, as well as Seiberg--Witten action,
are invariant under gauge transforms given by the formula:
$$
A\mapsto A+g^{-1}dg,\quad \Phi\mapsto g^{-1}\Phi
$$
where $g=e^{i\chi}\in C^\infty(X,\text{U}(1))$.

\textbf{Perturbed Seiberg--Witten equations}. In order to guarantee
the solvability of these equations we consider the perturbed
equations by plugging an appropriate self-dual 2-form
$\eta\in\Omega^2(X,i\mathbb R)$ into the second equation. As a
result we shall obtain the following equations
$$
\left\{
\begin{aligned}
D_A\Phi &=0\\
F_{A}^{+}+&\ \eta=(\Phi\otimes\Phi^{*})_0\,.
\end{aligned}\right.
$$

\subsection{Seiberg--Witten equations on 4-dimensional symplectic manifolds}
\label{ssec15}

\textbf{Dirac operator in symplectic case}. Let $X$ be a compact
symplectic 4-manifold provided with the symplectic form $\omega$ and
compatible almost complex structure $J$. Let $E\to X$ be a complex
Hermitian line bundle with a Hermitian connection $B$. We suppose
that $E$ is provided with the $\text{Spin}^c$-structure
$(W_E,\Gamma_E)$ and $\text{Spin}^c$-connection $\nabla_A$ where
$A\equiv A_E$ is the tensor product of the canonical
$\text{Spin}^c$-connection $A_{\text{can}}$ and $B$, determined by
the connection form
$$
A_E:=A_{\text{can}}\otimes\text{Id}+\text{Id}\otimes B.
$$
In this case the corresponding Dirac operator $D_A$ coincides with
$\bar\partial_B+\bar\partial_B^*$ and a section $\Phi\in
C^\infty(X,W_E^+)$ is given by the pair
$\Phi=(\varphi_0,\varphi_2)\in\Omega^0(X,E)\oplus\Omega^{0,2}(X,E)$.

\textbf{Seiberg--Witten equations in symplectic case}. The
complexified bundle $\Lambda^2_+\otimes\mathbb C$ of selfdual
2-forms in the considered case is decomposed into the direct sum of
subbundles
$$
\Lambda^2_+\otimes\mathbb C=\Lambda^{2,0}\oplus\mathbb
C[\omega]\oplus\Lambda^{0,2}.
$$
Respectively, the second SW-equation for the curvature decomposes
into the sum of three equations --- the one for the component,
parallel to $\omega$, the $(0,2)$-component and $(2,0)$-component.
The latter one is conjugate to the $(0,2)$-component and by this
reason is omitted below.

So the Seiberg--Witten equations take on the form
$$
\left\{
\begin{aligned}
\bar\partial_{B}\varphi_0 &+\bar\partial_{B}^{*}\varphi_2=0\\
F_{A_{\text{can}}}^{\omega} &+F_{B}^{\omega}=\frac{i}4
(|\varphi_0|^2-|\varphi_2|^2)-\eta^{\omega}\\
F_{B}^{0,2}+&\ \eta^{0,2}=\frac{\bar{\varphi}_0\varphi_2}{2}.
\end{aligned}\right.
$$

For the solvability of these equations we should impose some
topological condition on the first Chern class of the line bundle
$E$. Namely, we shall require that the following inequality is
satisfied (cf. \cite{Tau}):
\begin{equation}
\label{solv} 0\leq c_1(E)\cdot[\omega]\leq c_1(K)\cdot[\omega]
\end{equation}
where $K(X)=\Lambda^{*,0}(T^*X)$ is the canonical bundle of $X$ and
$[\omega]$ is the cohomology class of the form $\omega$.

\section{Abelian Higgs model}
\label{sec2}

\subsection{Ginzburg--Landau Lagrangian}
\label{ssec21}

Static $(2+1)$-dimensional Abelian Higgs model is governed by the
\textit{Ginzburg--Landau Lagrangian}, defined on the plane $\mathbb
R^2_{(x_1,x_2)}$ with coordinates $(x_1,x_2)$, having the form
$$
\mathcal L(A,\Phi)=|F_A|^2+|d_A\Phi|^2+\frac14(1-|\Phi|^2)^2
$$
where $A$ is a $\text{U}(1)$-connection on $\mathbb
R^2_{(x_1,x_2)}$, represented by the 1-form
$$
A=A_1dx_1+A_2dx_2
$$
with smooth pure imaginary coefficients. The curvature $F_A$ of this
connection is given by the 2-form
$$
F_A=dA=\sum_{i,j=1}^2
F_{ij}dx_i\wedge dx_j= 2F_{12}dx_1\wedge dx_2
$$
with coefficients $F_{ij}=\partial_i A_j-\partial_j A_i,\quad
\partial_j:=\partial/\partial x_j$.

The variable $\Phi$ is the scalar field, given by a smooth
complex-valued function $\Phi=\Phi_1+i\Phi_2$ on $\mathbb
R^2_{(x_1,x_2)}$. The covariant exterior derivative $d_A\Phi$ in the
second term of Ginzburg--Landau Lagrangian is given by the formula
$$
d_A\Phi=d\Phi+A\Phi=\sum_{i=1}^2(\partial_i+A_i)\Phi\,dx_i.
$$

The term $\frac14(1-|\Phi|^2)^2$ is the most important ingredient in
Ginzburg--Landau Lagrangian. It is responsible for the nonlinear
character of the ''self-interaction'' of the field $\Phi$. We
require that $|\Phi|\to 1$ for $|x|\to\infty$. In a neighborhood of
a zero of $\Phi$ the vector field $\vec v=\nabla\theta$ behaves like
the hydrodynamical vortex, by this reason solutions of the
considered model are also called \textit{vortices}.

\subsection{Vortices}
\label{ssec22}

\textbf{Definition}. The \textit{potential energy} of our model is
given by the integral of Ginzburg--Landau Lagrangian
$$
U(A,\Phi)=\frac12\int\mathcal
L(A,\Phi)\,d^2x.
$$
The condition $|\Phi|\to 1$ implies that the considered problem has
an integer-valued topological invariant $d$, given by the rotation
number of the map $\Phi$, sending the circles of sufficiently large
radius to the topological circles.

Mathematically, vortices are the pairs $(A,\Phi)$, minimizing the
potential energy $U(A,\Phi)<\infty$ in a given topological class,
fixed by the value of $d$. If $d>0$ (resp. $d<0$) such pairs are
called \textit{$d$-vortices} (resp. \textit{$|d|$-antivortices}).

\textbf{Gauge transforms}. The potential energy $U(A,\Phi)$ is
invariant under the \textit{gauge transforms}, given by the formula:
$$
A\mapsto A+id\chi,\quad \Phi\mapsto e^{-i\chi}\Phi
$$
where $\chi$ is an arbitrary smooth real-valued function on $\mathbb
R^2_{(x_1,x_2)}$.

The \textit{moduli space of $d$-vortices} is defined as the quotient
$$
\mathcal M_d=\frac{\{\text{$d$-vortices $(A,\Phi)$}\}}{\{\text{gauge
transforms}\}}
$$
described by the following theorem of Taubes.

\textbf{Taubes theorem}. Introduce the complex coordinate
$z=x_1+ix_2$ in the plane $\mathbb R^2_{(x_1,x_2)}$ identifying
$\mathbb R^2_{(x_1,x_2)}$ with the complex plane $\mathbb C_z$.

\begin{theorem}[Taubes cf. \cite{JT}]
\label{taubes} For any unordered collection $z_1,\dots,z_d$ of $d$
points on the complex plane $\mathbb C$, some of which may coincide,
there exists a unique (up to gauge transforms) $d$-vortex $(A,\Phi)$
such that the map $\Phi$ vanishes precisely at the points
$z_1,\dots,z_d$ with the same multiplicities as for the collection
$z_1,\dots,z_d$.

Moreover, any critical point $(A,\Phi)$ of the functional
$U(A,\Phi)<\infty$ with vortex number $d>0$ is gauge equivalent to
some $d$-vortex. In other words, all solutions of the
Euler--Lagrange equations for the functional $U(A,\Phi)$ with finite
energy are stable and have minimal energy in their topological
class.
\end{theorem}

\textbf{Description of moduli space of vortices}. The Taubes theorem
implies that the moduli space $\mathcal M_d$ of $d$-vortices may be
identified with the vector space $\mathbb C^d$ by associating with
the collection $z_1,\dots,z_d$ the monic polynomial, having its
zeros precisely at given points $z_1,\dots,z_d$ with given
multiplicities. The antivortices with $d<0$ admit an analogous
description.

This result has the following physical interpretation. Solutions of
the Euler--Lagrange equations for the functional $U(A,\Phi)$ consist
either of vortices, or antivortices. Our model cannot contain
simultaneously both vortices and antivortices --- such bound states
should ''annihilate'' before the system is transformed to the static
state.

\subsection{Dynamical Ginzburg--Landau equations}
\label{ssec13}

\textbf{Ginzburg--Landau action functional}. Now we switch on the
time in our model by adding the variable $x_0=t$. In this case the
Higgs field $\Phi=\Phi(t,x_1,x_2)$ is given by a smooth
complex-valued function on the space $\mathbb R^3_{(t,x_1,x_2)}$,
and the 1-form $A$ is replaced by the form $\mathcal
A=A_0dt+A_1dx_1+A_2dx_2$ with coefficients $A_\mu=A_\mu(t,x_1,x_2)$,
$\mu=0,1,2$, being smooth functions with pure imaginary values on
the space $\mathbb R^3_{(t,x_1,x_2)}$. Denote by $A^0=A_0dt$ the
time component of $\mathcal A$ and by $A=A_1dx_1+A_2dx_2$ its space
component.

The \textit{potential energy} of the system is given by the same
formula, as before, i.e. $U(\mathcal A,\Phi)=U(A,\Phi)$, while the
\textit{kinetic energy} has the form
$$
T(\mathcal
A,\Phi)=\frac12\int\left\{|F_{01}|^2+|F_{02}|^2+
|d_{A^0}\Phi|^2\right\}\,dx_1dx_2
$$
where $F_{0j}$, $j=1,2$, are defined in the same way, as before,
i.e. $F_{0j}=\partial_0 A_j-\partial_j A_0$, and
$d_{A^0}\Phi=d\Phi+A_0\,dt$.

The described dynamical model is governed by the
\textit{Ginzburg--Landau action} functional:
$$
S(\mathcal A,\Phi) =
\int_0^{T_0}\left(T(\mathcal A,\Phi) - U(\mathcal
A,\Phi)\right)\,dt,
$$
and the Euler--Lagrange equations for this functional, called
otherwise the \textit{Ginzburg--Landau equations} (briefly:
GL-equations) have the form
$$
\left\{\begin{aligned}
\partial_1F_{01} &+\partial_2F_{02} =
-i\,\text{Im}(\bar\Phi\nabla_{A,0}\Phi)\\
\partial_0F_{0j} &+ \sum_{k=1}^2\varepsilon_{jk}\partial_kF_{12} =
-i\,\text{Im}(\bar\Phi\nabla_{A,j}\Phi),\ j=1,2 \\
(\nabla_{A,0}^2 &-\nabla_{A,1}^2-\nabla_{A,2}^2)\Phi =
\frac12\Phi(1-|\Phi|^2),
\end{aligned}\right.
$$
where
$$
\nabla_{A,\mu}=\partial_{\mu}+A_{\mu},\ \mu=0,1,2,
$$
and $\varepsilon_{12}=-\varepsilon_{21}=1,\
\varepsilon_{11}=\varepsilon_{22}=0$. The first of these equations
is of constraint type while the last one is a nonlinear wave
equation.

\textbf{Dynamical gauge transforms}. The GL-equations, as well as
the action $S(\mathcal A,\Phi)$, are invariant under the
\textit{dynamical gauge transforms}, given by the same formula, as
in the static case:
$$
A_\mu\mapsto A_\mu+i\partial_\mu\chi,\quad \Phi\mapsto
e^{-i\chi}\Phi,\quad \mu=0,1,2,
$$
only now $\chi$ is a smooth real-valued function on $\mathbb
R^3_{(t,x_1,x_2)}$.

Our main goal is to describe solutions of the above GL-equations up
to dynamical gauge transforms. The quotient of the space of
dynamical solutions modulo gauge transforms is called the
\textit{moduli space of dynamical solutions}.

For the analysis of dynamical solutions it is convenient to choose
the gauge function $\chi$ so that the time component of the
potential will vanish, i.e. $A_0=0$ (temporal gauge). Note that
after imposing this condition we still have gauge freedom with
respect to static gauge transforms.

\textbf{Configuration space}. In the temporal gauge the dynamical
solution of the GL-equations may be considered as a trajectory of
the form $\gamma: t\mapsto [A(t),\Phi(t)]$ where $[A,\Phi]$ denotes
the gauge class of the pair $(A,\Phi)$ with respect to static gauge
transforms. This trajectory lies in the \textit{configuration space}
\smallskip
$$
\mathcal N_d=\frac{\{\text{$(A,\Phi)$ with $U(A,\Phi)<\infty$ and
vortex number $d$}\}}{\{\text{static gauge transforms}\}}
$$
\smallskip
which contains, in particular, the moduli space of $d$-vortices
$\mathcal M_d$.

The configuration space $\mathcal N_d$ may be thought of as a
\textit{canyon} the bottom of which is occupied by the moduli space
$\mathcal M_d$ of $d$-vortex solutions. Respectively, one may
consider a dynamical solution as the trajectory $\gamma(t)$ of a
small ball rolling along the walls of the canyon. The lower is the
kinetic energy of the ball, the closer lies its trajectory to the
bottom. Our ball may even hit the bottom of the canyon but cannot
stop there since, having a non-zero kinetic energy, it should assent
the canyon wall again.

\section{Adiabatic limit construction}
\label{sec3}

Here we present only a brief formulation of the adiabatic limit
construction and its properties for the Ginzburg--Landau and
Seiberg--Witten equations (for a more detailed exposition cf.
\cite{Ser} and \cite{Tau}).

\subsection{Adiabatic limit in Ginzburg--Landau equations}
\label{ssec31}

\textbf{Adiabatic equation}. Consider a family of dynamical
solutions $\gamma_\epsilon$ of GL-equations, depending on a
parameter $\epsilon>0$, with trajectories $\gamma_\epsilon:
t\longmapsto[A_\epsilon(t),\Phi_\epsilon(t)]$. Suppose that the
kinetic energy of these trajectories
$$
T(\gamma_\epsilon):=\int_0^{T_0} T(\gamma_\epsilon(t))dt
$$
tends to zero for $\epsilon\to0$, proportional to $\epsilon$. Then
in the limit $\epsilon\to0$ the trajectory $\gamma_\epsilon$
converts into a static solution, i.e. a point of $\mathcal M_d$.
However, if we introduce on $\gamma_\epsilon$ the \textit{''slow
time''} $\tau=\epsilon t$ and consider the limit of the ''rescaled''
trajectories $\gamma_\epsilon(\tau)$ for $\epsilon\to0$ then in this
limit we shall obtain a trajectory $\gamma_0$, lying in $\mathcal
M_d$. Of course, such trajectories cannot be solutions of the
original dynamical equations since any of their points is a static
solution. However, they describe approximately dynamic solutions
with small kinetic energy.

This procedure is called the \textit{adiabatic limit}. In this limit
the original dynamical equations reduce to the \textit{adiabatic
equation} whose solutions are called the \textit{adiabatic
trajectories}.

\textbf{Adiabatic trajectories}. The following theorem gives an
intrinsic description of adiabatic trajectories in terms of the
space $\mathcal M_d$.

\begin{theorem}

The kinetic energy functional generates a Riemannian metric on the
space $\mathcal M_d$, called the \textsl{kinetic} or
\textsl{T-metric}. The adiabatic trajectories $\gamma_0$ are the
geodesics of this metric.
\end{theorem}

\textbf{Adiabatic principle}. The idea of the approximate
description of ''slow'' dynamical solutions in terms of the moduli
space of static solutions was proposed on an heuristic level by
Manton \cite{Man} who postulated the following \textit{adiabatic
principle}: for any geodesic trajectory $\gamma_0$ on the moduli
space of $d$-vortices $\mathcal M_d$ it should exist a sequence
$\{\gamma_\epsilon\}$ of dynamical solutions, converging to
$\gamma_0$ in the adiabatic limit.

A rigorous mathematical formulation and the proof of this principle
were given recently by Palvelev in \cite{Pal} (cf. also
\cite{Pal-Ser}).

\textbf{Adiabatic correspondence}. The adiabatic principle reduces
the description of scattering of vortices in our model to the
description of geodesics on the moduli space of $d$-vortices
$\mathcal M_d$ in the kinetic metric, i.e. to the solution of Euler
geodesic equation on the space $\mathcal M_d$ provided with
$T$-metric.

In other words we have the following correspondence, established by
the adiabatic limit:
$$
\left\{\parbox{3cm}{solutions of
GL-equations}\right\}\longleftrightarrow
\left\{\parbox{8cm}{geodesics of the moduli space of vortices in
$T$-metric}\right\}
$$

\subsection{Adiabatic limit in SW-equations on 4-dimensional symplectic
manifolds} \label{ssec32}

\textbf{SW-equations with scale parameter}. In order to study the
adiabatic limit in SW-equations we plug the scale parameter into
them. For that we set the perturbation $\eta$ equal to
$\eta=-F_{A_{\text{can}}}^{+}+\pi i\lambda\omega$ where $\lambda$ is
the scale parameter and introduce the normalized sections
$\alpha:=\frac{\varphi_0}{\sqrt{\lambda}}$,
$\beta:=\frac{\varphi_2}{\sqrt{\lambda}}$. The perturbed
Seiberg--Witten equations will take the form
$$
\left\{
\begin{aligned}
\bar\partial_{B}\alpha &+\bar\partial_{B}^{*}\beta=0\\
\frac{4 i}{\lambda}F_{B}^{\omega} &=4\pi+|\beta|^2-|\alpha|^2\\
\frac2{\lambda}F_{B}^{0,2} &=\bar{\alpha}\beta.
\end{aligned}\right.
$$
and will be called briefly the
$\text{SW}_{\lambda}$-\textit{equations}.

\textbf{Apriori estimates}. Suppose that the necessary solvability
condition \eqref{solv} is satisfied and SW-invariant of $X$ does not
vanish\footnote{For the precise definition of SW-invariant cf.
\cite{Don}, \cite{Sal}, \cite{Wit}}. Then these equations have a
solution $(\alpha_\lambda,\beta_\lambda)$ for sufficiently large
$\lambda$. This solution has the following behavior for
$\lambda\to\infty$.

\begin{theorem}[Taubes \cite{Tau}]
\label{apriori}
\begin{enumerate}
\item[(1)] $|\alpha_\lambda|\to 1$ everywhere outside the set of zeros
$\alpha_\lambda^{-1}(0)$;
\item[(2)] $|\beta_\lambda|\to 0$ everywhere together with its
derivatives of the 1st order.
\end{enumerate}
\end{theorem}

\textbf{Taubes construction}. Denote by
$C_\lambda:=\alpha_\lambda^{-1}(0)$ the zero set of the section
$\alpha_\lambda$. Then the curves $C_\lambda$ converge in the sense
of currents to some \textit{pseudoholomorphic divisor} POincar\'e
dual to the Chern class $c_1(E)$, i.e. a chain $\sum d_kC_k$,
consisting of connected pseudoholomorphic curves $C_k$ taken with
miltiplicities $d_k$. Simultaneously, the original SW-equations
reduce to a family of static GL-equations defined in the complex
planes normal to the curves $C_k$. The chain $\sum m_kC_k$ may be
considered as a complex analogue of adiabatic trajectory in the
(2+1)-dimensional case.

Conversely, in order to reconstruct a solution of Seiberg--Witten
equations from the chain $\sum m_kC_k$, the family of vortex
solutions in normal planes should satisfy a nonlinear
$\bar\partial$-equation which may be considered as a \textit{complex
analogue of the Euler equation} for adiabatic geodesics with
''complex time''.

\textbf{Adiabatic correspondence}. Thus, in this case we have the
following correspondence, established by the adiabatic limit:
$$
\left\{\parbox{3cm}{solutions of
SW-equations}\right\}\longleftrightarrow
\left\{\parbox{8cm}{families of vortex solutions in normal planes of
pseudoholomorphic divisors}\right\}
$$

\end{document}